\documentclass[12pt]{article}
\usepackage[usenames]{color}
\usepackage{setspace}
\usepackage{graphicx}
\usepackage{amsmath}
\usepackage{amssymb}
\usepackage{amsthm}
\usepackage{multirow}
\numberwithin{equation}{section}

\newcommand{\del}{\partial}

\newcommand{\TeV}{\mbox{TeV}}

\newcommand{\Z}[1]{{\mathbb Z}_#1}
\newcommand{\abs}[1]{\left| #1 \right|}

\newcommand{\bequ}{\begin{equation}}
\newcommand{\eequ}{\end{equation}}
\newcommand{\beqn}{\begin{eqnarray}}
\newcommand{\eeqn}{\end{eqnarray}}
\newcommand{\bctr}{\begin{center}}
\newcommand{\ectr}{\end{center}}
\newcommand{\bit}{\begin{itemize}}
\newcommand{\eit}{\end{itemize}}
\newcommand{\Ls}{\left(}
\newcommand{\Rs}{\right)}

\newcommand{\Ll}{\left[}
\newcommand{\Rl}{\right]}

\newcommand{\half}{{\frac12}}
\def\e{{\textrm e}}
\def\del{\partial}
\def\half{{\frac12}}

\def\abs#1{{\left|{#1}\right|}}
\def\vev#1{\langle #1 \rangle}
\def\del{\partial}
\def\half{{\frac12}}

\def\abs#1{{\left|{#1}\right|}}
\def\vev#1{\langle #1 \rangle}

\def\del{\partial}
\def\dslash{\del\kern-0.55em\raise 0.14ex\hbox{/}}

\def\rough#1{\raise.3ex\hbox{$#1$\kern-.75em\lower1ex\hbox{$\sim$}}}

\newcommand{\PRD}[3]{{\it Phys. Rev.} {\bf D{#1}} (19{#3}) {#2}}

\newcommand{\PRDM}[3]{{\it Phys. Rev.} {\bf D{#1}} {#2} (20{#3})}
\newcommand{\PRL}[3]{{\it Phys. Rev. Lett.} {\bf {#1}} {#2} (19{#3})}

\newcommand{\NPB}[3]{{\it Nucl. Phys.} {\bf B{#1}} (19{#2}) {#3}}

\newcommand{\PLB}[3]{{\it Phys. Lett.} {\bf B{#1}} (19{#2}) {#3}}
\newcommand{\PLBM}[3]{{\it Phys. Lett.} {\bf B{#1}}, {#2} (20{#3})}

\newcommand{\PTPM}[3]{{\it Prog. Theor. Phys.} {\bf {#1}} (20{#3}) {#2}}
\newcommand{\PTEP}[3]{{\it PTEP} {\bf {#1}}(20{#2}){#3}}
\newcommand{\ANN}[3]{{\it Ann. Phys. (N.Y.)} {\bf {#1}}, {#2} (19{#3})}

\newcommand{\jhep}[3]{{\it JHEP} {\bf {#1}} (20{#2}) {#3}}

\begin{document}
\begin{flushright}
{\small KYUSHU-HET-166}\\%
\end{flushright}
\begin{center}
{\LARGE\bf Gauge Symmetry Breaking Patterns\\ in \\an $SU(5)$ Grand Gauge-Higgs Unification}
\vskip 1.4cm
{\large  
Kentaro Kojima$^{a,}$\footnote{E-mail: kojima@artsci.kyushu-u.ac.jp}, 
Kazunori Takenaga$^{b,}$\footnote{E-mail: takenaga@kumamoto-hsu.ac.jp}, 
and 
Toshifumi Yamashita$^{c,}$\footnote{E-mail: tyamashi@aichi-med-u.ac.jp}
}
\\
\vskip 1.0cm
{\it $^a$ Faculty of Arts and Science, Kyushu University, Fukuoka 819-0395, Japan\\
$^b$ Faculty of Health Science, Kumamoto Health Science University, Izumi-machi, Kumamoto 861-5598, Japan\\
$^c$ Department of Physics, Aichi Medical University, Nagakute 480-1195, Japan%
}\\
\vskip 1.5cm
\begin{abstract}
We study gauge symmetry breaking patterns of the five-dimensional $SU(5)$ grand gauge-Higgs 
unification  compactified on an orbifold $S^1/\Z2$ with the Hosotani mechanism 
in the framework of the diagonal embedding method. We find matter contents 
that lead to the $SU(3)\times SU(2)\times U(1)$ gauge symmetry on 
the global minimum of the effective potential 
and also present examples of matter content for which 
each regular subgroup of $SU(5)$ is realized as vacuum configuration. 
The finite temperature phase transitions for the models with the gauge symmetry of 
the standard model at zero temperature and also for supersymmetric models are studied. 
We show in a certain model with supersymmetry that the vacuum of the standard model 
selected dynamically before the inflation continues to stay there up to the present.
\end{abstract}
\end{center}
\vskip 1.0 cm
\newpage
%
\section{Introduction}
The grand unified theory (GUT)~\cite{gut}, in which quarks and leptons are treated in the same footing, 
is an attractive idea and has been investigated in various aspects since its proposal.
The gauge group of the standard model, $SU(3)\times SU(2)\times U(1)$ is embedded into a GUT
group, so that the gauge symmetry of the GUT must be broken down to the 
$SU(3)\times SU(2)\times U(1)$ by some mechanism. This is not understood well at the
present. One often assumes the Higgs mechanism to break the gauge symmetry. In fact, the prototype $SU(5)$
GUT~\cite{gut} introduces the Higgs scalar field belonging to the  adjoint representation 
under the $SU(5)$ by hand, and the scalar field is assumed to have the vacuum 
expectation value in such a way that the $SU(5)$ breaks down to the gauge group of the standard model.
There is not any evidence, however, that it should be the Higgs mechanism, and thus
it is important to seek the mechanism of the gauge symmetry breaking.

The GUT in a framework of the higher dimensional gauge theory has also been 
proposed and studied~\cite{orbifoldGUT, ghgut}, where the unified gauge symmetry is broken by the 
boundary conditions with respect to the compactified extra dimensions~\cite{orbifoldGUT}.
Meanwhile, one 
of the interesting points of the higher dimensional gauge theory
is that it can provide the mechanism of the gauge symmetry breaking which is different 
from the Higgs mechanism, that is, the Hosotani mechanism~\cite{hosotani}. 
In the Hosotani mechanism, component gauge fields for compactified directions, which behave like the 
Higgs scalar fields at low energy and are closely related with the Wilson line phases, develop 
the vacuum expectation values to induce the gauge symmetry breaking. 
An advantage of the Hosotani mechanism is that one can compute the effective potential 
for the Wilson line phases, with no need to introduce additional counterterms~\cite{hierarchy}, 
and the vacuum expectation values for the component gauge fields are
dynamically determined by minimizing it  once one fixes matter content of the theory. 
Then, one understands definite origin of the potential which induces the gauge symmetry breaking.

It should be noted that the periodic nature of the phases 
enables us to study the global structure of the 
effective potential even if we consider physics around the GUT and the Planck scale. 
This is in sharp contrast to the usual effective potential in the quantum field theory, where the 
potential is given by the polynomial of the scalar fields, so that one can study only the 
local structure of the potential near the origin. 
When the number of the order parameters in the effective potential increases,  it becomes 
difficult to minimize the effective potential analytically. In recent years, however, the environment 
of numerical study is dramatically improved, which makes easier to find the global minimum 
of the effective potential. In addition, the idea of GUTs with the Hosotani 
mechanism is a very attractive from phenomenological and theoretical points of view as 
discussed below, so that it should be investigated extensively.

When one tries to apply the Hosotani mechanism to break the $SU(5)$ gauge symmetries of 
GUT models, 
one immediately encounters a difficulty. 
Namely, the zero modes of the component gauge fields belonging to the adjoint 
representation under the gauge group that remains unbroken against the boundary conditions 
tend to be projected out by the boundary conditions in models with chiral fermions. 
Thus, in most GUT models with the Hosotani mechanism, the $SU(5)$ symmetry that contains 
the standard model gauge symmetry is broken by the boundary condition~\cite{ghgut}.
We have shown, however, in the paper~\cite{kty} that the difficulty is overcome 
by applying the so-called diagonal embedding method~\cite{diagonal}, which is invented in the 
context of string theory, to the higher dimensional
gauge theory. 
The method enables us to have GUT models with the Hosotani breaking of the $SU(5)$ gauge 
symmetry without contradicting with the chiral fermions and 
study the gauge symmetry breaking of the GUT. 
What is striking is that the effective potential obtained by the diagonal embedding method 
has basically 
the same form with the one obtained for the $S^1$ compactification. 
In other words, the diagonal embedding 
method can be seen as a way to introduce chiral fermions in $S^1$ compactification.

In this paper, we study the gauge symmetry breaking patterns of the models of the 
five-dimensional $SU(5)$ GUT 
compactified on an $S^1/\Z2$ by utilizing the diagonal embedding method. 
We refer to the models as the $SU(5)$ grand gauge-Higgs unification.
The $SU(5)$ gauge symmetry is broken by the Hosotani mechanism to one of its regular subgroups 
according to the matter content of the theory. 
We present examples of the matter content that realize the vacuum configuration 
with each regular subgroup of the $SU(5)$. 
We emphasize that matter contents, for which the
$SU(3)\times SU(2)\times U(1)$ gauge symmetry is realized on the global minimum of the theory, 
are found. One of the matter contents that realize the gauge symmetry of the standard model at the vacuum 
does not include the periodic fermions belonging to the adjoint representation, 
contrary to the expectation~\cite{congru}.

It has been known that the finite temperature field theory provides us a useful tool to study 
the phase transition, in the early Universe, {\it etc}~\cite{dj}. We also study the finite temperature phase transition 
for the matter contents that realize the gauge symmetry of the standard model at zero temperature 
in our $SU(5)$ models. 
We obtain the critical temperature and find the order of the phase transition. In the limit of 
high temperature, the 
fermions do not contribute to the effective potential, while the bosons do to yield
the symmetry restoration of the $SU(5)$ at a certain temperature (if light scalar fields are absent).

We apply the analysis of the finite temperature to the supersymmetric (SUSY) version of 
the $SU(5)$ grand gauge-Higgs unification. 
This scenario with the SUSY breaking scale being $O(1)$~TeV is phenomenologically 
very interesting in itself. 
For example, it generally predicts the existence of the light adjoint chiral supermultiplets 
with masses of the same scale and we may find a hint of the breaking of the GUT gauge symmetry
at the $\TeV$ scale experiments~\cite{gGHU-DTS, gGHU-pheno}. 
In addition, if a specific vacuum which is realized as a local minimum in Ref.~\cite{kty} is 
assumed, the so-called doublet-triplet splitting problem can be naturally solved~\cite{gGHU-DTS}.

In the SUSY version, one may also expect different behaviors of phase transitions 
from the non-SUSY cases because scalar fields are introduced in contrast to the latter. 
We find models of the SUSY $SU(5)$ grand gauge-Higgs unification in 
which the desired vacuum, where the doublet-triplet splitting problem can be naturally 
solved, is realized for the wide range of the temperature. 
The models may provide us an interesting possibility for the vacuum selection in the early 
Universe. Namely, the desired vacuum is dynamically selected 
in the epoch with very high temperature which may be exist
before the inflation
and continues to stay there up to the present.

This paper is organized as follows. In section $2$, after the brief introduction of the diagonal 
embedding method, we study the gauge symmetry breaking patterns
of the $SU(5)$ grand gauge-Higgs unification by the analyses of the one-loop effective 
potential. In section $3$, the finite temperature phase transition is studied for some models introduced 
in the section $2$. We also consider the supersymmetric version of the scenario and study 
the behavior at finite temperature and address the vacuum selection in the early Universe. 
The final section is devoted to conclusions and discussions.
We present some formulae necessary in the discussion of the section $3$ in the appendix.  
%
\section{Gauge symmetry breaking of an $SU(5)$ grand gauge-Higgs unification}
%
%
\subsection{Diagonal embedding method}
%
Let us first review quickly the diagonal embedding method 
which makes it possible for the Hosotani mechanism to break the $SU(5)$ gauge symmetry
consistent with chiral fermions. The simplest setting is a five-dimensional $SU(5)$ model on 
an orbifold $S^1/\Z2$ compactification with its radius being of the GUT scale. 
The detailed discussions are given in our previous paper~\cite{kty}, and readers 
who are familiar with the method can go to the next subsection.

In the simplest setting of the method, we prepare two copies of the gauge symmetry
which are exchanged by a discrete symmetry $\Z2$. We consider the theory 
with $SU(5)_1\times SU(5)_2\times \Z2$ symmetry, where the five-dimensional gauge field
for the $SU(5)_i~(i=1,2)$ is denoted by $A_M^{(i)}~(i=1,2)$. The $A_M^{(1)}$ and $A_M^{(2)}$
are related with each other by the $\Z2$. Here, $M=(\mu, y)=(0\mbox -3, 5)$ is a 
five-dimensional Lorentzian index, where the coordinate of the extra dimension is 
denoted by $y$ and the circumference of the $S^1$ is
$L=2\pi R$.

We impose the boundary conditions on the gauge fields at the two fixed 
points, $y_0=0, y_{\pi}=\pi R$ of the $S^1/\Z2$ as
\begin{equation}
A_{\mu}^{(1)}(y_i - y)=A_{\mu}^{(2)}(y_i + y),\qquad
A_y^{(1)}(y_i -y) =-A_y^{(2)}(y_i + y),
\label{shiki1}
\end{equation}
where we have used the notation $y_i (i=0, \pi)$.
We define the eigenstate under the operation of the $\Z2$ by
$X^{(\pm)}\equiv (X^{(1)}\pm X^{(2)})/\sqrt{2}$. Then, we see that 
$A_{\mu}^{(+)}$ and $A_y^{(-)}$ satisfy the Neumann boundary condition
at the both fixed points and thus have the zero modes. It implies that
the $SU(5)_1\times SU(5)_2$ breaks down to their diagonal part $SU(5)_{\rm diag}$
whose gauge field in the four-dimensional effective theory 
is $A_{\mu}^{(+)}$.  One says that our $SU(5)$ GUT symmetry
is embedded in the diagonal part. At the same time, one obtains the zero mode of the adjoint scalar field
$A_y^{(-)}$ under the $SU(5)_{\rm diag}$, and as shown in the previous paper, 
it composes the Wilson line phase,
\begin{equation}
W={\cal P}~{\rm exp}\biggl(
ig\int_0^{2\pi R}dy~~\frac1{\sqrt{2}}A_y^{(-)a}(T^a_1 -T^a_2)
\biggr),
\label{shiki2}
\end{equation}
where $\cal P$ stands for the path-ordered integral, $g$ is the common gauge coupling
constant, $T_i^a(i=1,2)$ is the generator of the $SU(5)_i (i=1,2)$ symmetry, and $a$ is an $SU(5)$
adjoint index. If we consider the fundamental representation ${\bf R_1}={\bf 5}, {\bf R_2}={\bf 1}$
for concreteness, we can parametrize the vacuum expectation value of $A_y^{(-)}$ as
\begin{equation}
\frac1{\sqrt{2}}gL\vev{A_y^{(-)a}T^a_{\bf 5}}
=2\pi~{\rm diag.} (\alpha_1, \alpha_2, \alpha_3, \alpha_4, \alpha_5)=2\theta^aT^a
\quad \mbox{with}\quad
\sum_{j=1}^5\alpha_j=0,
\label{shiki3}
\end{equation}
where we have used the remaining $SU(5)$ degrees of freedom to diagonalize the $A_y^{(-)a}T_{\bf 5}^a$.
Thus, the Wilson line phase is written as 
\begin{equation}
W={\rm exp}~\biggl(2\pi i ~{\rm diag.}\left(\alpha_1, \alpha_2, \alpha_3, \alpha_4, \alpha_5\right)\bigg).
\label{shiki4}
\end{equation}
We see that the order parameter $W$ is invariant under the shift $\alpha_j\to\alpha_j+1$.

Let us next introduce the fermion fields, $\Psi^{(1)}({\bf R}, {\bf 1})$ and 
$\Psi^{(2)}({\bf 1}, {\bf R})$, where
${\bf R}$ denotes a representation under the $SU(5)$ and are the $\Z2$ partner each other. 
We do not consider matter fields that are non singlet under the both gauge groups,  
$SU(5)_1$ and $SU(5)_2$, for simplicity. 
We impose the boundary condition,
\begin{equation}
\Psi^{(1)}(y_i -y) =\eta_{\Psi}^i\gamma^5\Psi^{(2)}(y_i +y),
\label{shiki5}
\end{equation}
where the parameter $\eta_{\Psi}^{i=0,\pi}$, which is associated with each fermion, takes
$+1$ or $-1$, and the product $\eta_{\Psi}^0\eta_{\Psi}^{\pi}$
gives the periodicity for the $S^1$ direction. The eigenstate $\Psi^{(\pm)}$ of the $\Z2$ 
has the zero mode for $\eta_{\Psi}^0\eta_{\Psi}^{\pi}=1$, while it does not  
for $\eta_{\Psi}^0\eta_{\Psi}^{\pi}=-1$.
Note that the zero modes are vector-like as $\Psi_L^{(\pm)}$ obeys the same boundary 
conditions as $\Psi_R^{(\mp)}$. The chiral 
fermions, for example the quarks and leptons in the standard model, can 
be put on the boundaries, though we do not consider their effects 
in this paper for simplicity.

As studied in the paper~\cite{kty}, the Kaluza-Klein (KK) mass spectrum for the $\Psi^{(\pm)}$ 
with $\eta_{\Psi}^0\eta_{\Psi}^{\pi}=1$ is given by
\begin{equation}
m_{\rm KK} ^{(n)}R=n+\frac{gR}{\sqrt{2}}\vev{A_y^{(-)a}} T^a_{\bf R} 
=n+\frac{\theta^a}{\pi}T^a_{\bf R},\quad n\in {\mathbb Z},
\label{shiki6}
\end{equation} 
and that for the $\Psi^{(\pm)}$ with $\eta_{\Psi}^0\eta_{\Psi}^{\pi}=-1$ by the same form 
in Eq.~(\ref{shiki6}) with replacement $n\to n+1/2$, which 
implies that the KK mass spectrum is basically the same form with the one obtained for
the $S^1$ compactification. Thus, each contribution from the $\Psi^{(+)}$ and $\Psi^{(-)}$ on the
orbifold $S^1/\Z2$ forms the same contribution from a fermion field on the $S^1$.
Then, once we fix the matter content whose representation is $\bf R$ under the $SU(5)$, one can
immediately write down the  contribution from the fermion to the effective potential by the help of the 
knowledge of the $S^1$ compactification. And one can study the breaking of the $SU(5)$ 
gauge symmetry through the Hosotani mechanism.

In non-SUSY models, we do not introduce the scalar fields in this paper, 
as they are generally expected to be superheavy due to the quantum corrections. 
In cases where the corrections are canceled to realize light scalars $\Phi^{(1)}({\bf R}, {\bf 1})$ and 
$\Phi^{(2)}({\bf 1}, {\bf R})$, by fine-tuning or the SUSY, we should impose the boundary condition, 
\begin{equation}
\Phi^{(1)}(y_i -y) =\eta_{\Phi}^i\Phi^{(2)}(y_i +y),
\label{shiki5-2}
\end{equation}
where the parameter $\eta_{\Phi}^{i=0,\pi}$ takes $+1$ or $-1$, and the KK mass spectrum for the 
$\Phi^{(\pm)}$ is the same as the one in Eq.~(\ref{shiki6}). 
%
\subsection{Matter content and gauge symmetry breaking pattern}
%
Let us study the gauge symmetry breaking patterns of the $SU(5)$ grand gauge-Higgs 
unification. We introduce fermions whose representations under the $SU(5)$ gauge group
are ${\bf 24, 5, 10}$ and ${\bf 15}$ \footnote{
We note that the fermions in the conjugate representation, {\it e.g.} ${\bf \bar 5}$, 
give the same contributions as those in the corresponding representation, since the 
five-dimensional models are vector-like.} 
and whose periodicities 
are $\eta_{\Psi}^0\eta_{\Psi}^{\pi}=+1~(-1)$ corresponding to the
(anti) periodic boundary condition. Following the standard prescription~\cite{hosotani} to 
calculate the one-loop effective potential in the background in Eq.~(\ref{shiki3}), we 
obtain the contribution of each degree of freedom in the aforementioned four representations to the 
potential as, up to the overall sign $(-1)^{f+1}$, 
\begin{eqnarray}
V_{24}(\alpha_i, z, \delta, f)&=&
C\sum_{i\neq  j=1}^5 F\Bigl(\alpha_i -  \alpha_j +\frac{\delta}{2},z,f\Bigr),
\label{shiki7}\\
V_{5}(\alpha_i, z, \delta, f)&=&
C\sum_{i=1}^5F\Bigl(\alpha_i+\frac{\delta}{2},z,f\Bigr),
\label{shiki8}\\
V_{10}(\alpha_i, z, \delta, f)&=&
C\sum_{1\leq i < j\leq 5}
F\Bigl(\alpha_i + \alpha_j +\frac{\delta}{2},z,f\Bigr),
\label{shiki9}\\
V_{15}(\alpha_i, z, \delta, f)&=&
C\sum_{1\leq i \leq  j\leq 5}~
F\Bigl(\alpha_i + \alpha_j +\frac{\delta}{2},z,f\Bigr),
\label{shiki10}
\end{eqnarray}
where we accommodate the temperature $T$ (normalized by $L$ as $z=LT$) dependence for later 
convenience, which should be set to zero in this section.
Here we have defined the function,
\begin{equation}
F(x,z=0,f)\equiv \sum_{w=1}^{\infty}\frac{1}{w^5}\cos(2\pi w x),
\label{shiki11}
\end{equation}
which represent{s} the contribution of each KK tower, 
and the overall constant $C$, which we shall ignore for the numerical analyses, is
\begin{equation}
C\equiv \frac{3}{4\pi^2}\frac{1}{L^5}.
\label{shiki12}
\end{equation}
The parameter $f$ stands for the fermion number and takes $f=1~(0)$ for fermions (bosons), and 
the fields satisfying the (anti) periodic boundary condition take $\delta=0~(1)$. 
We incorporate the suppressed overall sign $(-1)^{f+1}$ into the numbers of 
the degrees of freedom.  
Then, the total effective potential is given by
\begin{eqnarray}
V_{\rm eff}(\alpha_i,z)&=&
N_gV_{24}(\alpha_i, z, 0, 0)+4N_{24}^{(+)}V_{24}(\alpha_i, z, 0, 1)
+4N_{24}^{(-)}V_{24}(\alpha_i, z, 1, 1)\nonumber\\
&&
+4N_{5}^{(+)}V_{5}(\alpha_i, z, 0, 1)
+4N_{5}^{(-)}V_{5}(\alpha_i,, z, 1, 1)
+4N_{10}^{(+)}V_{10}(\alpha_i, z, 0, 1)\nonumber\\
&&
+4N_{10}^{(-)}V_{10}(\alpha_i, z, 1, 1)
+4N_{15}^{(+)}V_{15}(\alpha_i, z, 0, 1)
+4N_{15}^{(-)}V_{15}(\alpha_i, z, 1, 1), 
\label{shiki13}
\end{eqnarray}
where $N_g$ and the coefficients $4$ are the on-shell degrees of freedom for the gauge boson 
and fermion in the five dimensions, respectively.
In the non-SUSY case at zero temperature, $N_g=-3$ and $z=0$.
Here we have denoted the flavor numbers specifying the matter content as
\begin{equation}
(N_{24}^{(+)}, N_{24}^{(-)}, N_{5}^{(+)}, N_{5}^{(-)}, N_{10}^{(+)}, N_{10}^{(-)}, N_{15}^{(+)}, N_{15}^{(-)}).
\label{shiki14}
\end{equation}
The $(+), (-)$ denotes the periodicity of the field which corresponds to $\delta=0, 1$, respectively. 
The subscript number stands for the representation under the $SU(5)$ gauge group. 

\begin{table}[]
\begin{center}
\begin{tabular}{|c|c|c|c|c|c|c|c|c|}\hline
Case & $N_{24}^{(+)}$ &$N_{24}^{(-)}$ &$N_5^{(+)}$ &$N_5^{(-)}$ &$N_{10}^{(+)}$ &$N_{10}^{(-)}$ &$N_{15}^{(+)}$ &$N_{15}^{(-)}$ \\\hline\hline
1 &   1 & 0 & 0 & 0&   1&   0&   0&   1    \\\hline   
2 &   0 & 0 & 0 & 0&    7&   0&   0&   5     \\\hline   
3 &   1 & 0 & 0 & 0&    0&   0&   1&   1     \\\hline   
 4 &   1 & 0 & 1 & 0&    0&   1&   1&   0   \\\hline   \hline
5 &   0 & 0 &0  & 0& 0   &   0&   0& 0     \\\hline
6 &   0 & 0 & 1 & 1&    1&   0&   0&   3     \\\hline\hline
7  & 1 & 0 & 1 & 0&    0&   0&   0&   1    \\\hline 
8  & 1 & 0 & 1 & 0&    0&   0&   1&   0     \\\hline\hline
9  & 1 & 0 & 0 & 1&    0&   0&   1&   1     \\\hline\hline 
10 &   1 & 0 & 2 & 0&    0&   0&   0&   0     \\\hline\hline 
11  & 1 & 0 & 0 & 1&    1&   1&   0&   0    \\\hline \hline
12 &   1 & 0 & 0 & 0&    0&   0&   0&   0     \\\hline
\end{tabular}
\caption{Matter contents in the $SU(5)$ grand gauge-Higgs unification.}
\end{center}
\end{table}
\begin{table}[]
\begin{center}
\begin{tabular}{|c|c|c|c|c|c|c|}\hline
Case & $\alpha_1$ &$\alpha_2$ &$\alpha_3$ &$\alpha_4$ &$\alpha_5$ &Gauge symmetry\\\hline\hline
1 &   0.5 & 0.5 & 0 & 0& 0 &\multirow{4}{*}{$SU(3)\times SU(2)\times U(1)$}   \\\cline{1-6}
2 &   0.5 & 0.5 & 0 & 0&  0 & \\\cline{1-6}
 3 &  0.1 & 0.1 & 0.6 & 0.6 &  0.6  &\\\cline{1-6}
 4 & 0.223158  &0.223158  &0.223158  &0.665263 & 0.665263   &\\\hline   \hline
 5&   $k/5$&  $k/5$ &  $k/5$ &  $k/5$&    $k/5$ &\multirow{2}{*}{$SU(5)$}\\\cline{1-6}   
 6 &  0 &  0 &  0 &  0 &   0  &\\\hline  \hline
7&    0.5 &0.5  & 0.5  &0.5 &0 &  \multirow{2}{*}{$SU(4)\times U(1)$}\\\cline{1-6}
8&   0.287091& 0.287091 &  0.287091& {0.287091} & 0.851637  &\\\hline \hline
9&    0.139502 & 0.139502 & 0.139502  &0.679867  &  0.901628 &$ SU(3)\times U(1)^2$\\\hline\hline
10&   0.465989  &0.465989 &  0.534011&0.534011&0&$ SU(2)^2\times U(1)^2$\\\hline \hline
11&  0.154787 & 0.25036& 0.700439  & 0.947207& 0.947207 & $SU(2)\times U(1)^3$\\\hline \hline
12&   0.2& 0.4 & 0.6 & 0.8  &0  &$U(1)^4$\\\hline 
\end{tabular}
\caption{Vacuum expectation values and gauge symmetry breaking patterns for the cases 
given in Table $1$. The parameter $k$ takes $0, 1, 2, 3$ and $4$.
Here, we have shifted $\alpha_j$ into an interval $[0,1)$, instead of imposing the traceless 
condition which can be recovered by using the periodicity $\alpha_j\sim\alpha_j+1$.}
\end{center}
\end{table}

The parameter $\alpha_j$ is related with the Wilson line phase in Eq.~(\ref{shiki4}), 
so that the physical region is compact, 
reflecting its phase nature. Hence, one can 
study the global structure of the effective potential in Eq.~(\ref{shiki13}) 
even if we consider physics around the Planck scale. 
This is very contrast to the usual effective potential which is given by the
polynomial of the order parameters of the scalar fields and the analyses of 
the potential for the huge scale region around the Planck scale is unreliable. 
This is a great advantage of our approach to the breaking of the GUT gauge symmetry.

We show the matter contents we have chosen in the Table $1$. For each 
set of the matter content given in Table $1$,  we minimize the potential numerically to obtain 
the vacuum expectation values $\alpha_j$. Then, we understand how the $SU(5)$ is
broken down to its regular subgroup via the Hosotani mechanism,
by evaluating the Wilson line phase in Eq.~(\ref{shiki4}).
In particular, we are very much interested in the matter content that can realize 
the $SU(3)\times SU(2)\times U(1)$ gauge symmetry of the standard model
on the global minimum of the effective potential. 
We summarize our results in the Table $2$ in which the vacuum expectation values and 
the realized gauge symmetry on the global minimum are specified for each set of the matter content.

Let us note that the example of the matter 
content studied in~\cite{kty}, the case $6$, realizes the standard model gauge symmetry on a
local minimum of the effective potential. It may be 
worth to note here that the case $2$ realizes that the $SU(5)$ breaks down to 
the $SU(3)\times SU(2)\times U(1)$ even though it {does} 
not contain the fermions belonging to the ${\bf 24}$ 
representation, contrary to the understanding until now~\cite{congru}. We 
find examples  of matter contents, 
for which each regular subgroup
of $SU(5)$ is realized on the global minimum of the effective potential. 
%
\section{Grand gauge-Higgs unification at finite temperature}
%
In this section, we study the finite temperature phase transition for the models with the standard 
model gauge group on the global minimum studied in the previous section and for some SUSY models. 
At high temperature, the global minimum of the potential and vacuum configuration of 
the Wilson line phase would be altered due to existence of temperature dependent contribution to 
the effective potential. 
Thus, if the Universe has an epoch with very high temperature,
the finite temperature phase transitions may have implications for
grand unified models.

We follow the standard prescription, say, the Euclidean time formulation, to 
obtain the effective potential at finite temperature~\cite{dj}. The Euclidean time $\tau$ direction is, 
then, compactified on
$S^1_{\tau}$ whose circumstance is $1/T$, where $T$ is the temperature. 
We shall ignore the Wilson line 
degrees of freedom arising from the $S^1_{\tau}$, since no nontrivial vacuum configuration of 
the Wilson line phase is expected; the detailed discussion on the 
phase is given in Ref.~\cite{sakatake}. 
%
\subsection{Non supersymmetric models at finite temperature}
%
If we turn on the temperature $T$, the effective potential in one-loop approximation 
from the fermions belonging to the ${\bf 24}, {\bf 5}, {\bf 10}$ and ${\bf 15}$ representations 
are given by Eqs.~(\ref{shiki7})-(\ref{shiki10}) respectively, with the function for 
nonzero $z=LT$ which is treated as a free parameter in the theory, 
\begin{equation}
F(x, z, f)\equiv \sum_{\omega=1}^{\infty}{\frac1{\omega^5}}\cos(2\pi \omega x)
+2 {z^5} 
\sum_{\omega=1}^{\infty}\sum_{l=1}^{\infty}
\frac{(-1)^{fl}}{[(\omega z)^2 +l^2]^{\frac52}}\cos(2\pi \omega x),
\label{shiki15}
\end{equation}
as derived in Appendix~\ref{Sec:EffPotAtFT}.
The first (second) term in the right-hand side in Eq.~(\ref{shiki15}) 
is the zero (finite) temperature contribution.
Let us note that the (anti) periodicity of the bosons (fermions) for the Euclidean time direction 
manifestly appears as $(-1)^{fl}$ in the summation of the second term in 
Eq.~(\ref{shiki15}), which originally takes the form of $\cos(\pi fl)$ 
in Eq.~(\ref{Eq:EffPotAtFT}). The 
total effective potential is given by Eq.~(\ref{shiki13}), again with $N_g=-3$ but nonzero $z$.

We are interested in the behavior of the models at finite temperature
for the cases, $1,2,3$ and $4$ where the $SU(3)\times SU(2)\times U(1)$ is realized 
on the global minimum
at zero temperature. We numerically obtain the critical temperature $z_c$ for the four 
cases.\footnote{For the numerical analyses, we set the upper bound on the summation 
with respect to $\omega, l$ in Eq.~(\ref{shiki15}) to $100$. } 
Let us present the results for the cases, $1, 2, 3$ and $4$ in Tables $3, 4, 5$ and $6$, respectively. 
We write there the vacuum expectation values $\alpha_j$, the depths and 
the realized gauge symmetries for the global minimum and for some energetically lower local minima
at two temperatures near the critical temperature.

The critical temperatures for the cases $1,2$ and $3$ are commonly obtained as $z_c=0.720975$ 
by solving numerically the concrete form of an equation shown in Eq.~(\ref{Tc}) which is resulted by
the differences among the contributions in Eqs.~(\ref{shiki7})-(\ref{shiki10}) 
with the vacuum expectation values $\alpha_j$ corresponding to $G321, G41$ and $G5$. 
At this temperature, the three different vacua shown in the tables are 
degenerate with each other, and thus the phase transitions are of the 
first order.

The detailed discussion is given in Appendix~\ref{appendct}.
As for the case $4$, the phase transition occurs at $z=0.585788$ and is of the first order.
It can be shown that there is another phase transition at $z=0.830305$ above which 
the global minimum resides on $(0.4,0.4,0.4,0.4,0.4)$.

We confirm the expectation that the $SU(5)$ gauge symmetry is restored in the 
limit of high temperature because in the limit, the
fermions are decoupled from the system, so that the gauge bosons give dominant 
contributions to the effective potential~\cite{sakatake2}. 
\begin{table}[t]
\begin{center}
\begin{tabular}{|c|c||c|c|c|c|c|c|}\hline
$z=LT$&Depth &$\alpha_1$ &$\alpha_2$ &$\alpha_3$ &$\alpha_4$ &$\alpha_5$ &Gauge symmetry\\\hline\hline
      & $-19.2997$ &0.5 & 0.5 & 0 & 0& 0 &$G321$   \\\cline{2-8}
0.7209& $-19.2971$ &0.5 & 0.5 & 0.5 & 0.5&  0 &$G41$ \\\cline{2-8}
      & $-19.2917$ &0 & 0 & 0 & 0 &  0  &$G5$\\\hline  \hline 
      & $-19.3007$ & 0&  0 & 0 &  0&   0 &$G5$\\\cline{2-8}   
0.7210& $-19.2989$ & 0.5 &  0.5 &  0.5 &  0.5 &   0  &$G41$\\\cline{2-8}
      & $-19.2980$ &0.5 &0.5  & 0  &0 &0 &  $G321$\\\hline
\end{tabular}
\caption{Global and two local minima for the case $1$ near the 
critical temperature. 
Depths of the minima, which is 
values of the potential in Eq.~(\ref{shiki13}) normalized by overall factor $C$ 
in Eq.~(\ref{shiki12}), and values of $\alpha_i$ are shown.
$G321\equiv SU(3)\times SU(2)\times U(1), G41\equiv SU(4)\times U(1)$ and $G5\equiv SU(5)$.}
\end{center}
\end{table}
\begin{table}[t]
\begin{center}
\begin{tabular}{|c|c||c|c|c|c|c|c|}\hline
$z=LT$& Depth &$\alpha_1$ &$\alpha_2$ &$\alpha_3$ &$\alpha_4$ &$\alpha_5$ &Gauge symmetry\\\hline\hline
       & $-84.4617$ &0.5 & 0.5 & 0 & 0& 0 &$G321$   \\\cline{2-8}
0.7209 & $-84.4590$ &0.5 & 0.5 & 0.5 & 0.5&  0 &$G41$ \\\cline{2-8}  
       & $-84.4537$ &0 & 0 & 0 & 0 &  0  &$G5$\\\hline  \hline 
       & $-84.4595$ & 0&  0 & 0 &  0&   0 &$G5$\\\cline{2-8}   
0.7210 & $-84.4577$ & 0.5 &  0.5 &  0.5 &  0.5 &   0  &$G41$\\\cline{2-8} 
       & $-84.4568$ & 0.5 &0.5  & 0  &0 &0 &  $G321$\\\hline
\end{tabular}
\caption{Global and two local minima for the case $2$ near the 
critical temperature. Depths of the minima and values of $\alpha_i$ are shown.
$G321\equiv SU(3)\times SU(2)\times U(1), G41\equiv SU(4)\times U(1)$ and $G5\equiv SU(5)$.}
\end{center}
\end{table}
\begin{table}[t]
\begin{center}
\begin{tabular}{|c|c||c|c|c|c|c|c|}\hline
$z=LT$& Depth &$\alpha_1$ &$\alpha_2$ &$\alpha_3$ &$\alpha_4$ &$\alpha_5$ &Gauge symmetry\\\hline\hline
       & $-4.11202$ &0.1 & 0.1 & 0.6 & 0.6& 0.6 &$G321$   \\\cline{2-8}
0.7209 & $-4.10934$ &0.1 & 0.1 & 0.1 & 0.1&  0.6 &$G41$ \\\cline{2-8}   
       & $-4.10398$ &0.6 & 0.6 & 0.6 & 0.6 &  0.6  &$G5$\\\hline  \hline 
       & $-4.11335$ & 0.6&  0.6 & 0.6 &  0.6&   0.6 &$G5$\\\cline{2-8}   
0.7210 & $-4.11159$ & 0.1 &  0.1 &  0.1 &  0.1 &   0.6  &$G41$\\\cline{2-8}  
       & $-4.11070$ & 0.1 &0.1  & 0.6  &0.6 &0.6 &  $G321$\\\hline
\end{tabular}
\caption{Global and two local minima for the case $3$ near the 
critical temperature. Depths of the minima and values of $\alpha_i$ are shown.
$G321\equiv SU(3)\times SU(2)\times U(1), G41\equiv SU(4)\times U(1)$ and $G5\equiv SU(5)$.}
\end{center}
\end{table}
\begin{table}[t]
\begin{center}
\begin{tabular}{|c|c||c|c|c|c|c|c|}\hline
$z=LT$& Depth &$\alpha_1$ &$\alpha_2$ &$\alpha_3$ &$\alpha_4$ &$\alpha_5$ &Gauge symmetry\\\hline\hline
\multirow{2}{*}{0.5857} & $-19.2778$ &0.224824 & 0.224824 & 0.224824 & 0.662764 & 0.662764 &$G321$   \\\cline{2-8}
       & $-19.2759$ &0.311785 & 0.311785 & 0.311785 & 0.311785 & 0.752860 &$G41$ \\\hline   \hline 
\multirow{2}{*}{0.5858} & $-19.2773$ &0.311786 & 0.311786 & 0.311786 & 0.311786 & 0.752857 &$G41$\\\cline{2-8}   
       & $-19.2770$ &0.224826 & 0.224826 & 0.224826 & 0.662761 & 0.662761 &$G321$\\\hline  
\end{tabular}
\caption{Global and the energetically lowest local minimum for 
the case $4$ near the critical temperature. Depths of the minima and values of $\alpha_i$ are shown.
$G321\equiv SU(3)\times SU(2)\times U(1)$ and $G41\equiv SU(4)\times U(1)$.
}
\end{center}
\end{table}
%
\subsection{Supersymmetric models at finite temperature}
\label{Sec:SUSY-FT}
%
As stated in the introduction, the SUSY version of our $SU(5)$ model with the SUSY breaking scale 
$M_{\rm SB}$ being around the $\TeV$ scale is very attractive, especially on the desired vacuum
shown below in Eq.~(\ref{shiki18}), since the doublet-triplet splitting 
problem can be naturally solved there~\cite{gGHU-DTS}. 
In Refs.~\cite{gGHU-DTS,gGHU-pheno}, it is required that the desired vacuum is just one of 
minima, which is sufficient at zero temperature since the transition rate is extremely 
suppressed even if there are deeper vacua~\cite{tunnel}. The stability 
arises because the loop induced effective potential is suppressed by the SUSY 
breaking scale which is now much smaller than the compactification 
scale.\footnote{Hence, the mass scale of light adjoint chiral supermultiplets from zero-modes of $A_y$ 
and their superpartners is typically $O(M_{\rm SB})$ in this scenario.}
The smallness of the effective potential, at the same time, means that the stability of the vacuum 
can be easily affected by the finite temperature effects at high temperature $T\gg M_{\rm SB}$. 
Thus, it is important to study the finite temperature phase transition of the SUSY cases.

Because of the existence of the scalar fields, 
it is expected that the behavior of the effective potential should be modified from the non-SUSY 
version in which the only bosonic field is the gauge one.
We are, in particular, interested in the matter content that realizes the 
desired vacuum in Eq.~(\ref{shiki18}) as the global minimum for the wide range of the temperature.
We set the scale of the extra dimension $1/R$ around the GUT scale $M_{\rm GUT}$ as before. 
Then, $M_{\rm SB}$ is negligibly small compared with it. 
At high temperature $T\gg M_{\rm SB}$, we may neglect the effects of the 
zero-temperature SUSY breaking, and the SUSY breaking entirely comes from the finite 
temperature effect, say, the 
difference of the boundary conditions of bosons and fermions for the Euclidean time direction.
With this approximation, the zero temperature effective potential becomes vanishing.

The diagonal embedding method can be applied to supersymmetric models as well if we replace  
all the fields by the corresponding superfields. We again 
follow the standard prescription to calculate the finite temperature effective potential in
one-loop approximation in the background in Eq.~(\ref{shiki3}). 
Since we neglect the effects of the zero-temperature SUSY breaking, 
the contribution from the vectormultiplet is common with that from the adjoint hypermultiplet.  
The contributions from each degree of freedom of supermultiplets of ${\bf 24}, {\bf 5}, {\bf 10}$ 
and ${\bf 15}$ representations under the $SU(5)$
to the one-loop effective potential are given again by
Eqs.~(\ref{shiki7})-(\ref{shiki10}) respectively, but with replacing the function $F(x,z,f)$ by 
the SUSY version $F_S(x,z)$ defined as 
\begin{equation}
F_S(x, z)\equiv -F(x, z,0)+F(x, z,1)=-2z^5\sum_{\omega=1}^{\infty}
\sum_{l=1}^{\infty}\frac{\bigl(1-(-1)^l\bigr)}{\bigl[(\omega z)^2+l^2\bigr]^{\frac52}}\cos(2\pi \omega x).
\label{shiki16}
\end{equation}
The last variable of the contributions 
in Eqs.~(\ref{shiki7})-(\ref{shiki10}), the fermion number $f$, is dummy in the SUSY cases. 
Then, the total effective potential is given by Eq.~(\ref{shiki13}) with $N_g=4$.

Let us present an example of the matter content, 
for which the $SU(3)\times SU(2)\times U(1)$ gauge symmetry
is realized on the global minimum of the effective potential at finite temperature,
\begin{eqnarray}
(N_{24}^{(+)}, N_{24}^{(-)}, N_{5}^{(+)}, N_{5}^{(-)}, N_{10}^{(+)}, N_{10}^{(-)}, N_{15}^{(+)}, N_{15}^{(-)})
&=&(0, 1, 0, 0, 0, 1, 0, 0).
\label{shiki17}
\end{eqnarray}
We numerically find that for the above case, the 
phase transition occurs at $z_c\simeq 0.382008$, below which
the gauge symmetry of the standard model is realized on the global minimum whose vacuum 
expectation values are given by
\begin{equation}
(\alpha_1, \alpha_2,  \alpha_3,  \alpha_4,  \alpha_5)=(0.5, 0.5, 0, 0, 0).
\label{shiki18}
\end{equation}
As mentioned in the beginning of this subsection, we call it the desired vacuum. It may be 
worth to repeat that the vacuum is desirable
in the sense that it provides us an interesting solution to the doublet-triplet 
splitting problem  in certain SUSY models as studied in Ref.~\cite{gGHU-DTS}. 
Note that in the present case, the desired vacuum is obtained as 
the global minimum of the potential, not as its local minimum.

This model has the same nature of the finite temperature phase transition as 
the models without the supersymmetry. If the temperature increases further beyond 
the critical temperature, the gauge symmetry tends to be restored to 
the $SU(5)$ gauge symmetry for which the vacuum expectation values are given by
\begin{equation}
(\alpha_1, \alpha_2,  \alpha_3,  \alpha_4,  \alpha_5)=(0.2, 0.2, 0.2, 0.2, 0.2).
\label{shiki19}
\end{equation}
We also confirm the result in Eq.~(\ref{shiki19}) by minimizing numerically the 
effective potential consist of the bosonic degrees of freedom alone  
in the matter content in Eq.~(\ref{shiki17}) for the high temperature limit~\cite{sakatake2}. We 
summarize the results in Table {7} whose contents are the same as the ones in Tables {3,4,5 and 6}.
\begin{table}[t]
\begin{center}
\begin{tabular}{|c|c||c|c|c|c|c|c|}\hline
$z=LT$& Depth &$\alpha_1$ &$\alpha_2$ &$\alpha_3$ &$\alpha_4$ &$\alpha_5$ &Gauge symmetry\\\hline\hline
       &{$-2.77733$}&0.5 & 0.5 & 0 & 0& 0 &$G321$   \\\cline{2-8}
0.3820 &{$-2.77731$}&0.2 & 0.2 & 0.2 & 0.2&  0.2 &$G5$ \\\cline{2-8}   
       &{$-2.32086$}&0.5 & 0.5 & 0.5 & 0.5 &  0  &$G41$\\\hline  \hline
       &{$-2.77974$}&{0.2}&{0.2}&{0.2}&{0.2}&{0.2}&$G5$\\\cline{2-8}  
0.3821 &{$-2.77948$}& 0.5 &  0.5 &  0 &  0 &   0  &$G321$\\\cline{2-8}  
       &{$-2.32254$}& 0.5 &0.5  & 0.5  &0.5 &0 &  $G41$\\\hline
\end{tabular}
\caption{Global and two local minima for the matter content in Eq.~(\ref{shiki17}). 
Depths of the minima and values of $\alpha_i$ are shown.
$G321\equiv SU(3)\times SU(2)\times U(1), 
G41\equiv SU(4)\times U(1), G5\equiv SU(5)$.}
\end{center}
\end{table}

Let us next consider another model whose matter content is given by
\begin{eqnarray}
(N_{24}^{(+)}, N_{24}^{(-)}, N_{5}^{(+)}, N_{5}^{(-)}, N_{10}^{(+)}, N_{10}^{(-)}, N_{15}^{(+)}, N_{15}^{(-)})
&=&(0, 2, 1, 0, 0, 0, 0, 0).
\label{shiki20}
\end{eqnarray}
We numerically find that the $SU(3)\times SU(2)\times U(1)$ gauge symmetry persists as the 
global minimum of the effective potential for the wide range of the parameter $z < 1$ and $z \sim 10$
as far as our numerical analyses are concerned. 
The vacuum expectation values for both parameter regions are 
given by the desired vacuum in Eq.~(\ref{shiki18}). We also check that the desired vacuum is 
the global minimum of the effective potential consist of only the bosonic degrees of freedom 
in the matter content in Eq.~(\ref{shiki20}) for the high temperature limit~\cite{sakatake2}.

The above numerical analysis is performed in the range $0.1\leq z\leq10$, which corresponds to 
the temperature around the GUT scale. 
Although the reheating temperature can not be so high to avoid the so-called gravitino 
problem~\cite{gravitinoproblem}, the result obtained above may give us an interesting scenario 
of the vacuum selection in the evolution of the very early Universe before the inflation. 
Supposing that the Universe starts with very high temperature, say the Planck scale, 
there is an epoch with temperature corresponding to $z=O(1)$. Our 
result shows an expectation that the dynamics in the epoch naturally selects the desired vacuum.

Of course, in order to claim that the vacuum continues to stay there up to the present, we have to 
examine the potential for $z\ll 1$, where the one-loop effective potential is not 
reliable due to large logarithmic corrections 
as usual. For reliable analysis, such logarithmic terms should be resummed to result in the RG improved 
potential, which is not an easy task in the framework of the five dimensions. 
It is, however, instructive to see the behavior of the one-loop effective 
potential for $z\ll 1$ as shown below.
In this region, the {$w$-}dependence in the fraction in {Eq.}~%
(\ref{shiki16}) can be neglected for 
small $\omega$ $(\ll l/z)$, and then the function becomes 
\begin{equation}
F_S(x, z)\sim-2z^5\sum_{l}\frac{(1-(-1)^l)}{l^5}\sum_{\omega}\cos(2\pi \omega x),
\label{shiki21}
\end{equation}
aside form the overall factor. If we carried out the summation 
over $\omega$ to infinity, this function would become proportional 
to the periodic delta function, which vanishes except for the points $x=0$ (mod 1). 
Although this expression is not correct for large $\omega$, the contributions of the large $\omega$ tend 
to oscillate rapidly to cancel each other out approximately for $x\neq0$ (mod 1). 
Thus, we can see that it almost vanishes except for the neighborhoods of the points $x=0$ (mod 1). 
Noting that each KK tower that couples with $A_y$ gives a contribution expressed by the function 
$F_S(x, z)$ and the tower contains a massless mode on the points $x=0$ (mod 1), 
we come up with an anticipation that the effective potential is dominated only by the zero mode 
contribution. In fact, the decoupling theorem~\cite{decoupling} tells us that in the region away 
from the points $x=0$ (mod 1)
all the modes should be decoupled to give a vanishing contribution as they are much heavier than 
the temperature, which is consistent with the above behavior. 
This anticipation can be confirmed by comparing $F_S(0,z)$ with the zero mode contribution (and also 
their derivatives) in low temperature limit $z\ll1$. We present some formulae necessary in the
discussions given here in Appendix~\ref{Sec:EffPotAtFT}.

The above discussion justifies the evaluation of the the effective potential 
using the four-dimensional effective theory that consists only of the zero modes, 
as is the case with the Higgs quartic self coupling discussed in Ref.~\cite{EffTheo}.  
This fact is, of course, nothing but what is required in the field theory and thus may not be 
surprising at all, but it is quite nontrivial and amazing to see that it also holds in the 
nonrenormalizable five-dimensional models.

This consideration is not only academically interesting but also useful in the analysis of the 
effective potential for $z\ll1$, as it is much easier to study the RG improved 
effective potential in the four-dimensional effective theory. 
Although it appears to be difficult, at first glance, to carry out such a study model-independently 
even in the four dimensional effective theory, 
the SUSY allows us to derive a nontrivial conclusion.  
At (relatively) high temperature $M_{\rm SB}\ll T~(\ll1/R ~\sim M_{\rm GUT})$ where the zero temperature SUSY 
breaking is negligible, the interactions between the $A_y$ and the bosonic fields are determined 
by the gauge interaction and the three point interaction in the superpotential. 
It is well-known that the coupling constants of these interactions do not destabilize the system 
independently of their values, and thus of the RG running effects. 
At lower temperature $T\lesssim M_{\rm SB}$, the effective potential becomes dominated by the 
zero temperature contribution. 
This contribution is less controlled but we should impose that the desired vacuum is at 
least a local minimum if we would like to work there, and such parameters are chosen 
in Ref.~\cite{gGHU-pheno} actually.

In this way, we can see that the desired vacuum in Eq.~(\ref{shiki18}) does not become unstable 
for $z\ll1$, while it 
is not clear if it keeps being the global minimum as the global structure of the effective 
potential is hard to see in the effective-theoretical approach. 
Even if other vacua becomes deeper than the desired one, however, the transition rate is highly 
suppressed because the depth of the potential is very shallow compared with the distance 
between the minima, the order of $M_{\rm GUT}$, for $z\ll1$. 
Let us note that the condition $z\ll 1$ should always hold after the inflation in the history of the 
Universe. Thus, we conclude that in the models where the desired vacuum is the global minimum 
at $z\gtrsim1$ discussed above the vacuum is likely selected by the dynamics before the inflation 
and keeps staying there up to the present. 
\section{Conclusions and Discussions}
We have studied the gauge symmetry breaking patterns by the Hosotani mechanism 
of the models of the $SU(5)$ grand gauge-Higgs unification 
in the framework of the diagonal embedding method. The method 
enables us to have the zero mode for the adjoint scalar field and the chiral fermion simultaneously and to
utilize the one-loop effective potential obtained in the case of the $S^1$ 
compactification. We have investigated the vacuum structure of our models 
for the given matter contents through the analyses of the effective potential numerically.

We have found matter contents that realize the symmetry breaking down to 
each regular subgroup of the $SU(5)$. In 
particular, we have presented those that give us the desired vacuum 
with the $SU(3)\times SU(2)\times U(1)$ gauge symmetry as the global minimum. 
One of the matter contents that lead to 
the desired gauge symmetry breaking pattern 
does not include the periodic fermions belonging to the adjoint representation, 
contrary to the previous study~\cite{congru}.

We have also studied the finite temperature phase transition of the models of the 
$SU(5)$ grand gauge-Higgs unification with the $SU(3)\times SU(2)\times U(1)$ gauge 
symmetry being realized on the global minimum 
at zero temperature. We have numerically obtained the 
critical temperature and found that the phase transition is of the first order. 
The restoration of the $SU(5)$ gauge symmetry in
the high temperature limit is observed as it should.

Then, we examine the SUSY version of the $SU(5)$ grand gauge-Higgs unification at finite 
temperature. The SUSY version is phenomenologically fascinating when the SUSY breaking scale 
is set around the 
$\TeV$ scale as the existence of the light adjoint chiral supermultiplets with masses of the same 
scale are generally predicted to provide us a clue as to the GUT breaking 
accessible at the $\TeV$ scale experiments~\cite{gGHU-DTS,gGHU-pheno}. 
Furthermore, the doublet-triplet splitting problem can be naturally solved~\cite{gGHU-DTS} 
if the desired vacuum in Eq.~(\ref{shiki18}) is selected. 
Since the effective potential is suppressed by the SUSY breaking scale at zero temperature, 
the finite temperature 
effects are crucial to study if the desired vacuum is actually selected dynamically.

At enough high temperature, the SUSY breaking at zero temperature can be neglected and 
we have taken account of only the finite temperature effect for the SUSY breaking. 
We have analyzed the one loop effective potential to find a matter content for which the phase 
transition from the $SU(5)$ symmetric vacuum to the desired vacuum occurs at $z_c\sim 0.382$
as the temperature decreases. We have presented another matter content 
with which the desired vacuum stays as the global minimum even for $z\gtrsim1$.

We have also examined the effective potential in lower temperature 
region $M_{\rm SB}\ll T\ll1/R$ and found 
that its behavior is consistent with the decoupling theorem also in our nonrenormalizable 
five-dimensional setup.  
Though the effective potential without the RG improvement is not reliable in this region, 
the above observation shows that the effective potential is dominated by the zero mode contributions 
and thus justifies the use of the four-dimensional effective theory, as usual. 
Then, we have seen that the SUSY controls the finite temperature contribution well so that this 
contribution stabilizes the desired vacuum irrelevantly to the RG running. 
Thus, we conclude that in the models discussed in Sec.~\ref{Sec:SUSY-FT}, the vacuum selected
before the inflation keeps staying there up to the present. 
Given that in the model whose matter content is shown in Eq.~(\ref{shiki20}) the desired vacuum is 
the global minimum also for $z\gtrsim1$ and thus it is likely selected dynamically before the inflation, 
this may provide us an interesting scenario of the vacuum selection in the evolution of the early 
Universe.

Finally, we note that the examples analyzed in this paper are 
kind of toy models 
where we neglect the 
contributions from the matter sector of the standard model and do not discuss the gauge  coupling unification. 
When we construct a realistic model, we should redo our analysis. 
The present study creates an expectation that there exist several examples of appropriate matter 
contents.

%
%
%
\begin{center}
{\bf Acknowledgement}
\end{center}
The authors would like to thank Dr. Y. Ookouchi (Kyushu Uni{v}.) for valuable discussions.
%
%
\appendix
%
%
%
\section{Effective potential at finite temperature}
\label{Sec:EffPotAtFT}
%

Let us derive the function $F(x,z,f)$ in Eq.~(\ref{shiki15}) which represents the contribution of a 
KK tower with the mass spectrum 
\begin{equation}
 m_{\rm KK}^{(n)}=\frac{n+x+\delta/2}{R}. 
\end{equation}
Our space-time is 
\begin{equation}
S_{\tau}^1\times \mathbb R^3\times S^1/\Z2,
\end{equation}
where $S^1_{\tau}$ corresponds to the Euclidean time direction, for which the field satisfies
the definite boundary condition consistent with the quantum statistics. 
The $S^1/\Z2$ is the spacial extra dimension.

We evaluate the following quantity for the one-loop effective potential,
\begin{equation}
V=
(-1)^fN_{\rm deg}~\half  \frac{T}{L}\sum_{\bar l, n=-\infty}^{\infty}
\int \frac{d^{D-2}p_E}{(2\pi)^{D-2}}~{\rm ln}~
\Bigg[p_{E}^2 + (2\pi T)^2 \Bigl(\bar l +\frac f2\Bigr)^2+
\frac{(2\pi)^2}{(2\pi R)^2}\Bigl(n + x +\frac\delta 2\Bigr)^2\Biggr],
\end{equation}
where $f=0$ $(1)$ for bosons (fermions), $\delta=0$ $(1)$ for (anti)periodic fields with respect 
to the $S^1/\Z2$ direction.
Here, $T$, $L~(=2\pi R)$, $D=5$ and $p_E$  stand for the temperature, the circumference of the 
spacial extra dimension, the number of the dimensions and the Euclidean momentum, respectively. 
We assume the Wilson line phase for the $S_{\tau}^1$ direction is vanishing as suggested 
in Ref.~\cite{sakatake}, for simplicity.

Let us sketch the standard prescription to obtain the effective potential. Knowing that
\begin{equation}
 {\rm ln}~A=-\frac{d}{ds}A^{-s}\bigg|_{s\rightarrow 0} ~\mbox{and}~
A^{-s}=\frac{1}{\Gamma(s)}\int_0^{\infty}dt~t^{s-1}\e^{-At}
\stackrel{t=\frac{1}{X}}{=}\frac{1}{\Gamma(s)}\int_0^{\infty}dX~X^{-s-1}\e^{-\frac{A}{X}},
\end{equation}
we have
\begin{eqnarray}
V
=(-1)^fN_{\rm deg}\half \frac{T}{L}(-1)\sum_{\bar l,n=-\infty}^{\infty}
 \frac{\pi^{\frac{D-2}{2}}}{(2\pi)^{D-2}}
\int_0^{\infty}dt~t^{-\frac{D}{2}}
~\e^{-\Bigl((2\pi T)^2(\bar l +\frac f2)^2+ \frac{(2\pi)^2}{(2\pi R)^2}(n +x+\frac{\delta}{2})^2\Bigr)t}.
\end{eqnarray}
We employ the Poisson's formula for the modes, $\bar l$ and $n$,
\begin{eqnarray} 
\sum_{\bar l=-\infty}^{\infty}\e^{-(2\pi T)^2(\bar l +\frac f2)^2~t}&=&
\sum_{l=-\infty}^{\infty}\frac{1}{2\pi T}\sqrt{\frac{\pi}{t}}\e^{-\frac{1}{4t}(\frac{l}{T})^2-2\pi i l \frac f2},
\label{Poisson-l}
\\
\sum_{n=-\infty}^{\infty}\e^{-\frac{(2\pi)^2}{(2\pi R)^2}(n +x+\frac{\delta}{2})^2~t}&=&
\sum_{w=-\infty}^{\infty}R\sqrt{\frac{\pi}{t}}\e^{-\frac{1}{4t}(2\pi R w)^2-2\pi i w (x+\frac{\delta}{2})}.
\end{eqnarray}
We decompose the summation of $l$ into $l=0$ and $l\neq 0$ parts. The $l=0~(\neq 0)$ corresponds to the
zero (finite) temperature part. The $w=0$ part in the $t$ integration of the $l=0$ part 
which is independent of $x$ yields the UV divergence.
We just subtract the divergence, and we finally have, setting $D=5$, 
\begin{eqnarray}
V&=&(-1)^{f+1}N_{\rm deg}~
\frac{\Gamma(\frac{5}{2})}{\pi^{\frac52}L^5}
\Biggl[
\sum_{w=1}^{\infty}\frac{1}{w^5}\cos\Bigl[2\pi w\Bigl(x+\frac{\delta}{2}\Bigr)\Bigr]
\nonumber\\&&\hspace{35mm}
 +2(LT)^5 \sum_{l,w=1}^{\infty}\frac{\cos\Ls\pi lf\Rs}{\Ll(wLT)^2+l^2\Rl^{\frac52}}
\cos\Bigl[2\pi w\Ls x+\frac\delta2\Rs\Bigr]
\Bigg].
\label{Eq:EffPotAtFT}
\end{eqnarray}
The factor $\frac{\Gamma({\frac52})}{\pi^{\frac52}L^5}$ coincides with the overall constant $C$ 
defined in Eq.~(\ref{shiki12}), and the quantity in the parenthesis in the above right hand side 
reduces to the function $F(x,z,f)$ in Eq.~(\ref{shiki15}), by shifting $x+\delta/2\to x$.
%
\subsection{Zero mode contribution}
%
Next, let us evaluate the contribution only by the zero mode for the comparison with the 
low temperature region ($z\ll1$) of the $l\neq0$ part discussed in Sec.~\ref{Sec:SUSY-FT}.

In this calculation we set $n=0$ instead of taking the summation over $n$. 
Thus, we shall apply the Poisson's formula only for $\bar l$ in Eq.~(\ref{Poisson-l}), 
and then the integration over $t$ results in the modified Bessel function $K_2(x)$ as 
\begin{equation}
\int_0^{\infty}dt~t^{-3}
~\e^{-\frac{1}{4t}(\frac lT)^2 + \frac{(2\pi)^2}{(2\pi R)^2}x^2t}
=8\frac{T^2}{l^2}\frac{(2\pi)^2}{L^2}x^2
 K_2\Ls\abs{\frac{2\pi lx}{LT}}\Rs,  
\end{equation}
where we have made the shift $x+\delta/2\to x$ for simplicity. 
Taking away the overall factors, we see that the contribution to the function $F_S(x,z)$ 
defined in Eq.~(\ref{shiki21}) for the SUSY cases from the zero mode, $F_S^{(0)}(x,z)$, 
becomes\footnote{This is consistent with the result obtained in~\cite{sakatake2}.}
\begin{equation}
 F_S^{(0)}(x,z) = -\frac{2z^4}{3}\sum_{l=1}^{\infty}\frac{1-(-1)^l}{l^4}
              \Ls\frac{2\pi lx}{z}\Rs^2K_2\Ls\abs{\frac{2\pi lx}{z}}\Rs. 
\end{equation}
In low temperature limit $z\to0$, the value and the curvature of $F_S(x,z)$ at $x=0$, where the 
zero mode becomes massless, 
are analytically calculated by replacing the summation over $w$ by an integral as 
\begin{eqnarray}
 F_S(0,z)
  \to -2z^4\sum_{l=1}^\infty\int_0^\infty\frac{1-(-1)^l}{[s^2+l^2]^\frac52}ds
  = -\frac{4z^4}{3}\sum_{l=1}^{\infty}\frac{1-(-1)^l}{l^4}=-\frac{\pi^4}{36}z^4, \\
 \frac12\frac{d^2F_S}{dx^2}(0,z)
  \to (2\pi)^2z^2\sum_{l=1}^\infty\int_0^\infty\frac{1-(-1)^l}{[s^2+l^2]^\frac52}s^2ds
  =\frac{(2\pi)^2z^2}{3}\sum_{l=1}^{\infty}\frac{1-(-1)^l}{l^2}=\frac{\pi^4}{3}z^2.
\end{eqnarray}
Expanding the modified Bessel function $K_2(x)=\frac2{x^2}-\frac12+{\cal O}(x^2)$, we see that these 
coincide respectively with those of $F_S^{(0)}(x,z)$. 
This coincidence can be understood in a more general context by reminding that the Poisson's formula 
relates a function $f(x)$ with its Fourier transform $\hat f(k)$ as 
\begin{equation}
 \sum_{n=-\infty}^\infty f(n+x) = \sum_{w=-\infty}^\infty \hat f(w) \e^{2\pi iwx}. 
\end{equation}
When we replace the summation over $w$ by an integral at $x=0$, the right hand side becomes 
\begin{equation}
 \sum_{w=-\infty}^\infty \hat f(w) \e^{2\pi iwx}
 \quad\to\quad \int_{-\infty}^\infty \hat f(s) ds = f(0). 
\end{equation}

\section{Critical temperature}
\label{appendct}
%
%
In this appendix, we comment on the phase transition in the cases 1,2,3, for which the 
critical temperature is common, and at the temperature three vacua degenerate. 
%
\subsection{Cases 1,2}
%
As shown in the Tables $3$ and $4$ for the cases $1$ and $2$, respectively, the critical 
temperature is commonly given by $z_c=0.720975$ (and also for the  case 3 discussed next).  We
shall explain below that around $z_c$ three vacua whose vacuum expectation values are given by 
\begin{eqnarray}
\alpha_{0}&=&(0,0,0,0,0),\\
\alpha_{321}&=& (2,2,2,-3,-3)/2\sim(0,0,0,1,1)/2,\\
\alpha_{41}&= &(-4,1,1,1,1)/2\sim(0,1,1,1,1)/2, 
\end{eqnarray}
tend to be degenerate. Here we have taken account of the periodicity of $\alpha_j$.

For this purpose, we calculate the differences of the contributions from each degree of freedom 
in the representation ${\bf R=5, 10, 15,24}$ to the configurations $\alpha_{321}$ and $\alpha_{41}$ 
compared with those to $\alpha_0$. 
The differences 
$\Delta_{\bf R}(\alpha_i,z,\delta,f)=
(V_{\bf R}(\alpha_i,z,\delta,f)-V_{\bf R}(\alpha_0,z,\delta,f))/C$
are written as 
\begin{eqnarray}
&& \Delta_{\bf 5}(\alpha_{321},z,\delta,f)=\frac12\Delta_{\bf 5}(\alpha_{41},z,\delta,f)
 =2\Ls F\Ls\frac12+\frac\delta2,z,f\Rs-F\Ls\frac\delta2,z,f\Rs\Rs 
\nonumber\\&&\hspace{64mm}
 =2(-1)^\delta\Delta_f,\\
&& \Delta_{\bf 10}(\alpha_{321},z,\delta,f)=\frac32\Delta_{\bf 10}(\alpha_{41},z,\delta,f)
 =6(-1)^\delta\Delta_f,\\
&& \Delta_{\bf 15}(\alpha_{321},z,\delta,f)=\frac32\Delta_{\bf 15}(\alpha_{41},z,\delta,f)
 =6(-1)^\delta\Delta_f,\\
&& \Delta_{\bf 24}(\alpha_{321},z,\delta,f)=\frac32\Delta_{\bf 24}(\alpha_{41},z,\delta,f)
 =12(-1)^\delta\Delta_f,\\ 
\notag 
&& \Delta_f = F\Ls\frac12,z,f\Rs-F\Ls0,z,f\Rs
 \\&&\hspace{7mm}
= \sum_{w=1}^\infty\Ls\frac{(-1)^w-1}{w^5}
   +2z^5\sum_{l=1}^\infty\frac{(-1)^{fl}\Ls(-1)^w-1\Rs}{\Ll(wz)^2+l^2\Rl^{5/2}}\Rs,
\end{eqnarray}
where the periodicity of the function $F(x)$ defined in Eq.~(\ref{shiki15}), $F(x+1)=F(x)$, is used.

We see that the differences flip the sign when the periodicity $\delta=0,1$ is changed, 
and when there are no fields belonging to the {\bf5} representation the difference of the total 
effective potential between the points $\alpha_{321}$ and $\alpha_0$ is always $3/2$ times 
larger than that between $\alpha_{41}$ and $\alpha_0$. 
The latter explains that the three vacua degenerate at a certain temperature in cases with no 
{\bf5} fields. 
In such cases, the three vacua degenerate at the temperature determined by the equation 
\begin{equation}
 -3\times12\Delta_0
 +4\times6\Ls2N_{24}^{(+)}-2N_{24}^{(-)}+N_{10}^{(+)}-N_{10}^{(-)}+N_{15}^{(+)}-N_{15}^{(-)}
          \Rs\Delta_1
 =0, 
\end{equation}
which reduces to
\begin{equation}
 12\Ls-3\Delta_0+4\Delta_1\Rs =0, 
\end{equation}
commonly in the cases 1 and 2, whose concrete form is given as 
\begin{equation}
 12\Ls-\frac{31}{16}\zeta_R(5)
  +2z^5\sum_{w,l=1}^{\infty}\frac{(-1)^w-1}{\Ll (wz)^2+l^2\Rl^{5/2}}\Ls-3+4(-1)^l\Rs\Rs =0.
\label{Tc}
\end{equation}
Here, $\zeta_R(x)$ is the Riemann's zeta function. 
%
\subsection{case 3}
%
In the case 3, the three vacua that degenerate at the temperature are those shifted by the center 
of the $SU(5)$, $\alpha_c=(2,2,2,2,2)/5$ (mod $1$) from those in the previous cases, 
\begin{equation}
\alpha_{0}'=\alpha_{0}+\alpha_{c},\qquad
\alpha_{321}'=\alpha_{321}+\alpha_{c},\qquad
\alpha_{41}'=\alpha_{41}+\alpha_{c}. 
\end{equation}
The differences in this case 
$\Delta'_{\bf R}(\alpha_i',z,\delta,f)=
(V_{\bf R}(\alpha_i',z,\delta,f)-V_{\bf R}(\alpha_0',z,\delta,f))/C$
are 
\begin{eqnarray}
&& \Delta'_{\bf 5}(\alpha_{321}',z,\delta,f)=\frac12\Delta_{\bf 5}(\alpha_{41}',z,\delta,f)
 =2(-1)^\delta\Delta_f^{(1)},\\
&& \Delta'_{\bf 10}(\alpha_{321}',z,\delta,f)=\frac32\Delta_{\bf 10}(\alpha_{41}',z,\delta,f)
 =6(-1)^\delta\Delta_f^{(2)},\\
&& \Delta'_{\bf 15}(\alpha_{321}',z,\delta,f)=\frac32\Delta_{\bf 15}(\alpha_{41}',z,\delta,f)
 =6(-1)^\delta\Delta_f^{(2)},\\
&& \Delta'_{\bf 24}(\alpha_{321}',z,\delta,f)=\frac32\Delta_{\bf 24}(\alpha_{41}',z,\delta,f)
 =12(-1)^\delta\Delta_f^{(0)},\\ 
&& \Delta_f^{(q)} = F\Ls\frac12+q\times\frac25,z,f\Rs-F\Ls q\times\frac25,z,f\Rs,
\end{eqnarray}
where the parameter $q$ denotes the charge of the center of the $SU(5)$ group, $\Z5$.
By a similar discussion as before, we see that the three vacua degenerate at a certain temperature, 
and the temperature is determined again by Eq.~(\ref{Tc}) since the dependence on $\Delta_f^{(2)}$ 
is canceled in the case 3 and $\Delta_f^{(0)}=\Delta_f$.



\begin{thebibliography}{99}
%
%
%
\bibitem{gut} H. Georgi and S. L. Glashow, \PRL{32}{438}{74}.
%
%
\bibitem{orbifoldGUT} 
Y. Kawamura, \PTPM{103}{613}{00}; ibid {\bf 105} (2001) 691; ibid {\bf 105} (2001) 999;
L. J. Hall and Y. Nomura, \PRDM{84}{055003}{01}; ibid {\bf 65} (2002) 125012; ibid {\bf 66} (2002) 075004.
%
%
\bibitem{ghgut} 
  G.~Burdman and Y.~Nomura,
  Nucl.\ Phys.\ B {\bf 656}, 3 (2003);
%
  N.~Haba, Y.~Hosotani, Y.~Kawamura and T.~Yamashita,
  Phys.\ Rev.\ D {\bf 70}, 015010 (2004);
%
C. S. Lim and N. Maru, \PLBM{653} {320}{07}, Y. Hosotani and N. Yamatsu, \PTEP{2015}{15}{111B01}.
%
%
\bibitem{hosotani} Y. Hosotani, \PLB{126}{83}{309}; ibid {\bf 129}, 193 (1983), \ANN{190}{233}{89}.
%
%
\bibitem{hierarchy}
N.~V.~Krasnikov,
Phys.\ Lett.\ B {\bf 273}, 246 (1991);
%
H.~Hatanaka, T.~Inami and C.~S.~Lim,
Mod.\ Phys.\ Lett.\ A {\bf 13}, 2601 (1998);
%
%
 N.~Maru and T.~Yamashita,
 Nucl.\ Phys.\  B {\bf 754}, 127 (2006);
%
  Y.~Hosotani, N.~Maru, K.~Takenaga and T.~Yamashita,
  Prog.\ Theor.\ Phys.\  {\bf 118}, 1053 (2007).
%
%
\bibitem{kty} K. Kojima, K. Takenaga and T. Yamashita, \PRDM{84}{051701}{11}.
%
%
\bibitem{diagonal}  K. R. Dienes and J. March-Russell,  \NPB{479}{96}{113};
%
%
D. C.Lewellen, \NPB{337}{90}{61}; G. Aldazabal, A.~Font, L. E. Ibanez and A. M.Uranga,
\NPB{452}{95}{3}; J. Erler, \NPB{475}{96}{597}; Z. Kakushadze and S. H. H. Tye, \PRD{55}{7878}{97};
M. Ito {\it et al.}, \PRDM{83}{091703}{11}; \jhep{1112}{11}{100}.
%
%
\bibitem{congru} A. T. Davies and McLachlan,  \PLB{200}{88}{305}, \NPB{{317}}{89}{237}.
%
%
\bibitem{dj} L. Dolan and R. Jackiw, \PRD{9}{3320}{74}, S. Weinberg, \PRD{9}{3357}{74}.
%
%
\bibitem{gGHU-DTS} T. Yamashita, \PRDM{84}{115016}{11}.
%
%
\bibitem{gGHU-pheno} M. Kakizaki, S. Kanemura, H. Taniguchi and T. Yamashita, \PRDM{89}{075013}{14}.
%
%
%
%
\bibitem{sakatake} M. Sakamoto and K. Takenaga, \PRDM{76}{085016}{07}.
%
%
\bibitem{sakatake2} M. Sakamoto and K. Takenaga, \PRDM{80}{085016}{09}.
%
%
\bibitem{tunnel} M. J. Duncan and L. G. Jensen, \PLB{291}{92}{109}.
%
%
%
%
\bibitem{gravitinoproblem}  M. Y. Khlopov and A. D. Linde, \PLB{138}{84}{265}, 
J. R. Ellis, D. V. Nanopoulos and S. Sarkar, \NPB{259}{85}{175}.
%
%
\bibitem{decoupling}
  T.~Appelquist and J.~Carazzone,
  Phys.\ Rev.\ D {\bf 11}, 2856 (1975).
%
%
\bibitem{EffTheo} N. Haba, S. Matsumoto, N. Okada and T. Yamashita, \jhep{0602}{06}{073}.
%
%
\end{thebibliography}
\end{document}